%% file: paper.tex
\documentclass[runningheads,a4paper,floatfix]{llncs}

\include{preamble}
\urldef{\mailsa}\path|rubenm@andrew.cmu.edu|
\urldef{\mailsb}\path|sbjoshi@iith.ac.in|
\urldef{\mailsc}\path|{vasco.manquinho, ines.lynce}@tecnico.ulisboa.pt|
\urldef{\mailsd}\path|ines.lynce@tecnico.ulisboa.pt|

\begin{document}

\mainmatter

%\title{Contribution to Silver Jubilee of CP}
%\title{Comments to ``Incremental Cardinality Constraints for MaxSAT'' from CP 2014}
\title{Reflections on ``Incremental Cardinality Constraints for MaxSAT''}

%The published version of this article can be found at :
%\url{http://dx.doi.org/10.1007/978-3-319-19249-9_19}}

%\titlerunning{Silver Jubilee of CP}

\author{Ruben Martins\inst{1} \and Saurabh Joshi\inst{2} \and Vasco Manquinho\inst{3} \and In{\^e}s Lynce\inst{3}}

%\authorrunning{}

\institute{ Department of Computer Science, Carnegie Mellon University, United States \\
\mailsa 
\and 
Department of Computer Science and Engineering, IIT Hyderabad, India \\
\mailsb 
\and 
INESC-ID/Instituto Superior T{\'e}cnico, Universidade de Lisboa, Portugal \\
\mailsc
}

\toctitle{}
\tocauthor{}
\maketitle
%\input{abstract}
%\nocite{*}
\begin{abstract}
To celebrate the first 25 years of the International Conference on Principles and Practice of Constraint Programming (CP) the editors invited the authors of the most cited paper of each year to write a commentary on their paper. This report describes our reflections on the CP 2014 paper ``Incremental Cardinality Constraints for MaxSAT'' and its impact on the Maximum Satisfiability community and beyond.
\end{abstract}

\input{intro}
\input{origin}

\input{lookingback}

\input{impact}

\paragraph{Acknowledgement.} We would like to thank Naveen Pai for a preliminary study on the impact of cores when building incremental cardinality constraints for Independent Studies in Computer Science at Carnegie Mellon University~\cite{pai-poster}.

\bibliographystyle{splncs03}
\bibliography{biblio,cp14}

\end{document}

%% file: preamble.tex
\usepackage{subcaption}
\usepackage{amsmath}
\usepackage{amssymb}
\usepackage{listings}
\usepackage{multirow}
\usepackage{fancyvrb}
\usepackage{xcolor}

\usepackage{xspace}
\usepackage{tikz}
\usepackage{cite}
\usepackage{pgfplots}
 \usepackage{hyperref}

\input{refs}

\newcommand{\inroundb}[1]{\ensuremath{ \left( #1 \right)}}

\newcommand{\IT}[1]{\textit{#1}}
\newcommand{\TT}[1]{\texttt{#1}}

\newcommand{\MT}[1]{\ensuremath{\mathtt{#1}}}

\newcommand{\EM}[1]{\emph{#1}}
\newcommand{\Omit}[1]{}

\newcommand{\true}{$true$}

\definecolor{highlighter}{rgb}{0.85,0.85,0.85}

\makeatletter
\newbox\sf@box
{\def\sf@one{#1}%
\def\sf@two{#2}%
\setbox\sf@box\hbox
\bgroup}%
{ \egroup
\ifx\@empty\sf@two\@empty\relax
\def\sf@two{\@empty}
\fi
\ifx\@empty\sf@one\@empty\relax
\subfloat[\sf@two]{\box\sf@box}%
\else
\subfloat[\sf@one][\sf@two]{\box\sf@box}%
\fi}
\makeatother

\usepackage[linesnumbered,ruled,vlined]{algorithm2e}

\newif\ifhenabled

%% file: refs.tex
\newcommand{\figref}[1]{Fig.~\ref{Fi:#1}}

\renewcommand{\eqref}[1]{Eqn.~(\ref{Eq:#1})}

\newcommand{\tableref}[1]{Tab.~\ref{Ta:#1}}

%% file: intro.tex
\section{Introduction\label{Se:intro}}

During the first decade of the century, advances in Propositional
Satisfiability (SAT) algorithms resulted in the proposal of solving Maximum
Satisfiability (MaxSAT) using iterative calls to a SAT solver.
Previously, most MaxSAT solvers used a branch and bound approach enhanced
with lower bounding procedures and MaxSAT inference rules~\cite{manya-handbook09}.

However, the ability of SAT solvers to provide an unsatisfiable subformula
whenever an unsatisfiable call is made enabled the proposal of new algorithms for
MaxSAT~\cite{FM06,manquinho-sat09,msuncore-aaai11,morgado-constraints13}.
These new algorithms were orders of magnitude faster when solving several
sets of real-world problem instances.

Meanwhile, the incremental usage of SAT solvers~\cite{strichman-charme01,minisat-assumptions,nadel-sat12,audemard-sat13}
had provided significant gains in some domains where SAT algorithms
were iteratively being used.
However, incrementality had not yet been fully exploited in MaxSAT solving,
except for algorithms using a Sat-Unsat approach.

In Unsat-Sat algorithms for MaxSAT, at each iteration, a new instance
of the SAT solver was created and the formula was rebuilt from scratch.
As a result, almost all the knowledge from the previous iteration was lost.
The main reason for rebuilding the SAT solver is that some constraints
from previous iterations are no longer valid. Moreover, removing these
constraints would not be enough, as learned constraints would also have
to be removed. This cleaning process is not easy to perform efficiently.

\input{msu3}

Algorithm~\ref{alg:msu3} presents the MSU3 algorithm for partial MaxSAT.
Observe that, at each iteration, the cardinality constraint in line~\ref{msu3:satcall}
from one iteration is not valid to the next iteration. The set of literals might
have changed and the lower bound $\lambda$ is incremented.

In our paper~\cite{DBLP:conf/cp/MartinsJML14}, we proposed three techniques
that allow Unsat-Sat MaxSAT algorithms (such as Algorithm~\ref{alg:msu3})
to use the same SAT solver between iterations.
In particular, we propose the usage of (i) incremental blocking,
(ii) incremental weakening, and (iii) iterative encoding of cardinality constraints.
Note that rebuilding the SAT solver, includes having to rebuild the CNF
encoding of cardinality constraints. However, the iterative encoding of cardinality
constraints allows to gradually encode the cardinality constraint into CNF,
maintaining the auxiliary variables already used in the previous iteration.
Moreover, as a result of using incrementality, the internal state of the SAT
solver is maintained, as well as learned clauses discovered in the previous
iterations. Experimental results from our paper and subsequent MaxSAT
evaluations clearly show the effectiveness of the proposed techniques.
We also note that current state of the art MaxSAT solvers extensively use
the incremental techniques originally proposed in the paper.

The remainder of the paper is organized as follows. Section~\ref{Se:origin}
describes the problem that motivated this approach and its application
to MaxSAT. Next, we study and discuss the effectiveness of these incremental
techniques. Finally, in section~\ref{Se:impact}, the impact of the proposed
ideas on subsequent MaxSAT technology is revised.

%% file: msu3.tex
\DontPrintSemicolon
\SetKwFunction{soft}{soft}
\SetKwFunction{SAT}{SAT}
\SetKwFunction{decomposeSoft}{partitionSoft}
\SetKwFunction{first}{first}
\SetKwFunction{weight}{weight}
\SetKwFunction{encodeCNF}{CNF}
\SetKwFunction{min}{min}
\SetKwData{result}{satisfiable assignment to}
\SetKwData{unsat}{UNSAT}
\SetKwData{sat}{SAT}
\SetKwData{minc}{min$_\textnormal{c}$}
\SetKwData{true}{true}
\SetKwData{st}{st}
\SetVlineSkip{1pt}
\begin{algorithm}[!t]
  \small
  \KwIn{$\varphi = \varphi_h \cup \varphi_s$}
  \KwOut{satisfying assignment to $\varphi$}
  %$(\mathcal A, \mathcal U) \gets (\emptyset, \emptyset)$\;
  %$(\st, \nu, \varphi_C) \gets \SAT(\varphi_h, \mathcal A)$\tcp*[r]{\footnotesize check if the MaxSAT formula is \unsat}
  %\If{$\st =$ \unsat}{
  %  \Return{\unsat}
  %}
  $(\varphi_W, V_R, \lambda) \gets (\varphi, \emptyset, 0)$\;
  \While{\true}{
    $(\st, \nu, \varphi_C) \gets \SAT(\varphi_W \cup \{ \encodeCNF( \sum_{r\in V_R} r \leq \lambda ) \})$ \label{msu3:satcall}\;
    \If{$\st = \sat$}{
      \Return{$\nu$}\tcp*[r]{\footnotesize satisfying assignment to $\varphi$}
    }
    \ForEach{$c \in (\varphi_C ~\cap~ \varphi_s)$}{
       $V_R \gets V_R \cup \{ r \}$\tcp*[r]{\footnotesize r is a new variable}
       $c_R \gets c \vee r$\tcp*[r]{\footnotesize clause $c$ was not previously relaxed}
       $\varphi_W \gets (\varphi_W \setminus \{ c \}) \cup \{ c_R \}$\;
    }
    $\lambda \gets \lambda + 1$\;
  }
  \caption{MSU3 Algorithm for partial MaxSAT}\label{alg:msu3}
\end{algorithm}

%% file: origin.tex
\section{Origin} \label{Se:origin}

\begin{figure}[t]
\begin{center}
%\begin{tabular}{c@{\qquad}c@{\qquad}c}
\begin{tabular}{c@{\qquad}c}
%\begin{minipage}{1.8in}

%\begin{subfigure}{\linewidth}
	\subcaptionbox{\label{Fi:tsoreorder}}[.28\linewidth]
	{
\centering
\begin{scriptsize}
\begin{tabular}{ccc}
\multicolumn{3}{c}{$\MT{x=0,y=0;}$} \\
& & \\
\begin{minipage}{0.55in}
%\centering
%\begin{array}{ccl}
$s_1$ :  \MT{x=1;} \\
$s_2$  :  \MT{r1=y;} \\
%\end{array} \\
\end{minipage} & \large{$\parallel$} &
\begin{minipage}{0.55in}
%\begin{array}{ccl}
$s_3$  :  \MT{y=1;} \\
$s_4$  :  \MT{r2=x;} \\
%\end{array} \\
\end{minipage} \\
& & \\
\multicolumn{3}{c}{\MT{assert(r1==1||r2==1);}}\\
\end{tabular}
\end{scriptsize}
%\caption{}
%\caption{Reordering in \TT{x86}}
%\label{Fi:tsoreorder}
}
&
%\end{subfigure}
%\end{minipage} &

%\begin{minipage}{1.8in}
%\begin{subfigure}{\linewidth}

%\end{subfigure}
%\end{minipage} &
%\end{figure}
%\begin{minipage}{2.8in}
%\begin{subfigure}{\linewidth}
\subcaptionbox{\label{Fi:innocent}}[.25\linewidth]
{
%\centering
\begin{scriptsize}
\begin{tabular}{ccc}
\multicolumn{3}{c}{$\MT{x=0,y=0,w=0,z=0;}$} \\
& & \\
\begin{minipage}{0.55in}
\centering
%\begin{array}{ccl}
$s_1$  :  \MT{z=1;} \\
$s_2$  :  \MT{p1=w;} \\
$s_3$ :  \MT{x=1;} \\
$s_4$  :  \MT{r1=y;} \\
%\end{array} \\
\end{minipage} & \large{$\parallel$} &
\begin{minipage}{0.55in}
%\begin{array}{ccl}
$s_5$  :  \MT{w=1;} \\
$s_6$  :  \MT{p2=z;} \\
$s_7$  :  \MT{y=1;} \\
$s_8$  :  \MT{r2=x;} \\
%\end{array} \\
\end{minipage} \\
& & \\
\multicolumn{3}{c}{\MT{assert(r1==1||r2==1);}} \\
\multicolumn{3}{c}{ \MT{assert(p1+p2>=0);}}\\
\end{tabular}
\end{scriptsize}
%\caption{}
%\caption{Program with \IT{innocent} and {culprit} pairs}
%\label{Fi:innocent}
}
\end{tabular}
%\end{subfigure}
%\end{minipage} \\
%\end{tabular}
\end{center}
\caption{(\subref{Fi:tsoreorder}) Reordering in TSO. 
%(\subref{Fi:psoreorder}) Reordering in PSO.
(\subref{Fi:innocent}) A program with \IT{innocent} and \IT{culprit} reorderings }
\end{figure}

The motivation for incremental cardinality constraints \cite{DBLP:conf/cp/MartinsJML14}
comes from a problem of a completely different domain of automated program repair for weak memory models.

Modern multicore CPUs implement optimizations such as \IT{store buffers} and
\IT{invalidate queues}.  These features result in weaker memory consistency
guarantees than sequential consistency (SC)~\cite{lamportsc}.  Though such
hardware optimizations offer better performance, the weaker consistency has
the drawback of intricate and subtle semantics, thus making it harder for
programmers to anticipate how their program might behave when running on such
architectures.  For example, a pair of statements can appear to have been
executed out of the program order.

Consider the program given in \figref{tsoreorder}.  Here, \TT{x} and \TT{y}
are shared variables whereas \TT{r1} and \TT{r2} are thread-local variables or registers. 
Statements $s_1$ and $s_3$ perform write operations.  Because of store
buffering, these write operations may not be reflected immediately in the memory. 
Next, both threads may proceed to perform the read operations $s_2$ and
$s_4$.  Since the write operations might still not have hit the memory,
stale values for \TT{x} and \TT{y} may be read in \TT{r2} and \TT{r1},
respectively.  This will cause the assertion to fail.  Such behavior is
possible with architectures that implement \IT{Total Store Order (TSO)},
which allows write-read reordering.  Note that on an hypothetical
architecture that guarantees sequential consistency, this would never
happen.  However, due to store buffering, a global observer might witness
that the statements are executed in the order $(s_2,s_4,s_1,s_3)$, which
results in the assertion failure.  We say that $\inroundb{s_1,s_2}$ and
$\inroundb{s_3,s_4}$ have been reordered.

Since these reorderings are non-deterministic, architectures usually provide
\EM{fence} (or \EM{memory barrier}) instructions to allow a programmer to restrict such
reorderings. In this case, a fence between $\inroundb{s_1,s_2}$ and $\inroundb{s_3,s_4}$ is needed to ensure
that the assertion does not fail.

We distinguish approaches that aim to restore sequential consistency (SC)
and approaches that aim to ensure that a user-provided assertion holds. 
Since every fence incurs in a performance penalty, it is desirable to keep the
number of fences to a minimum.  

Consider the example given in \figref{innocent}. Here, \TT{x,y,z,w} are
shared variables initialized to $0$.  All other variables are thread-local. 
A processor that implements total store ordering (TSO) permits a read of a
global variable to precede a write operation to a different global variable
when there are no dependencies between the two statements.
Note that if $(s_3,s_4)$ or $(s_7,s_8)$ is reordered, the assertion will be violated.
We shall call such pairs of statements \IT{culprit pairs}. By contrast, the pairs
$(s_1,s_2)$ and $(s_5,s_6)$ do not lead to an assertion violation
irrespective of the order in which their statements execute.  We shall
call such pairs of statements \IT{innocent pairs}.  A tool that restores SC would insert
four fences, one for each pair mentioned earlier.  However, only two fences
(between $s_3,s_4$ and $s_7,s_8$) are necessary to avoid the assertion
violation.

There are approaches which look at a counterexample provided by a model checker for such programs and tries to avoid some minimal set of reorderings in order to avoid assertion violations. For most of these approaches, a large number of innocent pairs is a bane and would result in a lot of expensive queries to the underlying model checker.
Reorder Bounded Model Checking (ROBMC)~\cite{fm15} addresses this issue by exploring only those behaviors where the number of behaviors is bounded by some parameter $k$.
For every pair of statements $s_i,s_j$ which can potentially get reordered, a new variable $a_{ij}$ is introduced such that a reordering is allowed only if $a_{ij}$ is \true. 
Then, a cardinality constraint $a_{ij} \leq k$ enforces the bound on reordering. In~\cite{fm15}, it is shown that ROBMC results in much lesser number of queries to the model checker, thus making it much more efficient. Initially, $k=1$ so that all the counterexamples with only one culprit pair can be eliminated.

In general, there could be assertion violations, which will be triggered only if
more than $1$ culprit pairs are reordered. Therefore, for soundness, after making the program safe
for $k=1$, the bound $k$ must be increased to a higher value to check if there are counter examples
for this higher bound. Note that to check for assertion violation at a higher bound $k'$, only
constraint that is required to change is from $\left( \sum a_{ij} \leq k \right)$
to $\left( \sum a_{ij} \leq k' \right)$.
The rest of the formula which encodes all possible
program behaviors remains the same. We move to a higher bound $k'$ only when
at a lower bound $k$ the program is declared safe, which is usually indicated by the corresponding
formula representing program behaviors being unsatisfiable. This is where, \textit{incremental cardinality constraints} play a crucial role by allowing us to increase the upper bound of the cardinality constraints when approaching from an unsatisfiable region.

%% file: lookingback.tex
\section{Looking Back\label{Se:lookingback}}

The experimental evaluation conducted in the CP'14 paper~\cite{DBLP:conf/cp/MartinsJML14} showed a clear improvement when using incremental cardinality encodings. Not only the number of solved instances increased but, on average, the incremental version was $3.6 \times$ faster than the corresponding non-incremental version. However, Ansotegui et al.~\cite{DBLP:journals/ai/AnsoteguiG17} stated that the improvement may not be due to incrementality but rather to the way the cardinality encoding is built. Note that, in the incremental approach, the cardinality encoding is built taking into consideration the structure of the unsatisfiable subformulas (cores) found by the algorithm. 

Looking back to our original experiments, we did not fully explore the reason behind the improvements. In this section, we perform additional experiments that show that our original insights were correct and that the \textit{primary reason for the performance gain is incrementality} and not the way the cardinality encoding is built. However, the way the cardinality encoding is built may also have a small benefit for the performance of the solver and should be the target of further study.

We restrict ourselves to the non-incremental and incremental versions of the MSU3 algorithm since this was the most efficient algorithm presented in our CP'14 paper~\cite{DBLP:conf/cp/MartinsJML14}. To clarify the nature of the improvement, we ran five variants described in \tableref{msu3} and implemented on top of the \texttt{open-wbo} framework~\cite{martins-sat14}. Column ``Incremental'' indicates if the SAT solver is reused between iterations, column ``Encoding Structure'' indicates if the cardinality constraint is rebuilt from scratch in each iteration and column ``Reuse Cores'' indicates if the cardinality constraint is built using the Unsat cores found in the incremental version.

More specifically:
Version (ni) corresponds to the classic implementation of the MSU3 algorithm with no incrementality where the SAT solver is not reused and the cardinality constraint is rebuilt in each iteration; Version (i) corresponds to the fully incremental version proposed in the paper; Version (ni-c) does not reuses the SAT solver between iterations, but the cardinality constraint is non-incrementally rebuilt using the unsatisfiable cores from the incremental version (i); Version (ni-s) also does not reuses the SAT solver, but the cardinality constraint is built according to the structure of the unsatisfiable cores found by the algorithm (i.e. the structure of the cardinality encoding follows the structure of the cores found); Version (ni-s-c) also does not reuses the SAT solver, but the cardinality constraint is built according to the structure of the unsatisfiable cores found by algorithm (i).
The goal of testing all these variants is to clarify if the improvement is coming due to the way the encoding is built or due to the incrementality of the approach. All experiments were run on StarExec~\cite{starexec} using Intel Xeon E5-2609 processors (2.40GHz) with a memory limit of 32GB and time limit of 1,800 seconds. We used the same benchmarks as in our CP'14 paper, which corresponds to the 627 partial industrial MaxSAT instances from the MaxSAT Evaluation 2013.

\begin{table}[!t]
\begin{center}
\begin{tabular}{|l|r|r|r|}
\hline
Version & Incremental & Encoding Structure & Reuse Cores\\\hline
ni     & No  & Complete Rebuild             & No\\
ni-c   & No  & Complete Rebuild             & From incremental version\\
ni-s   & No  & Structure of the Unsat cores & No\\
ni-s-c & No  & Structure of the Unsat cores & From incremental version\\
i      & Yes & Structure of the Unsat cores & No\\
\hline
\end{tabular}
\end{center}
\caption{Different versions of MSU3}\label{Ta:msu3}
\end{table}

\begin{figure}
\centering
\includegraphics[scale=0.85]{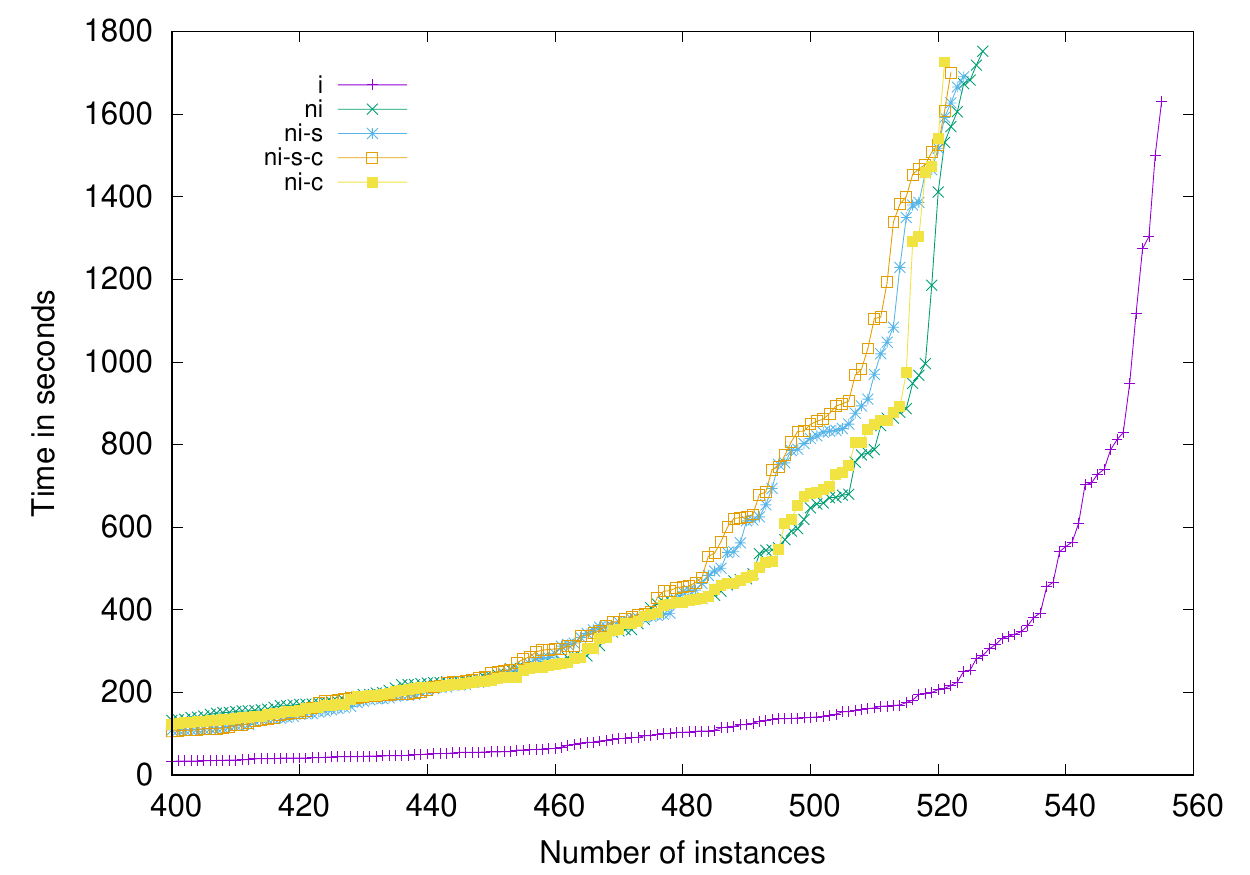}
\caption{Running times of the different non-incremental and incremental versions}\label{Fi:versions}
\end{figure}

\figref{versions} shows a cactus plot with the running times of the different non-incremental and incremental versions. We can see that the incremental version clearly outperforms the non-incremental versions and it is the main reason for the performance of the incremental MSU3 algorithm. On the other hand, all non-incremental versions solve a similar number of instances.

\begin{figure}[!t] 
  \begin{subfigure}[b]{0.5\linewidth}
    \centering
    \includegraphics[width=0.75\linewidth]{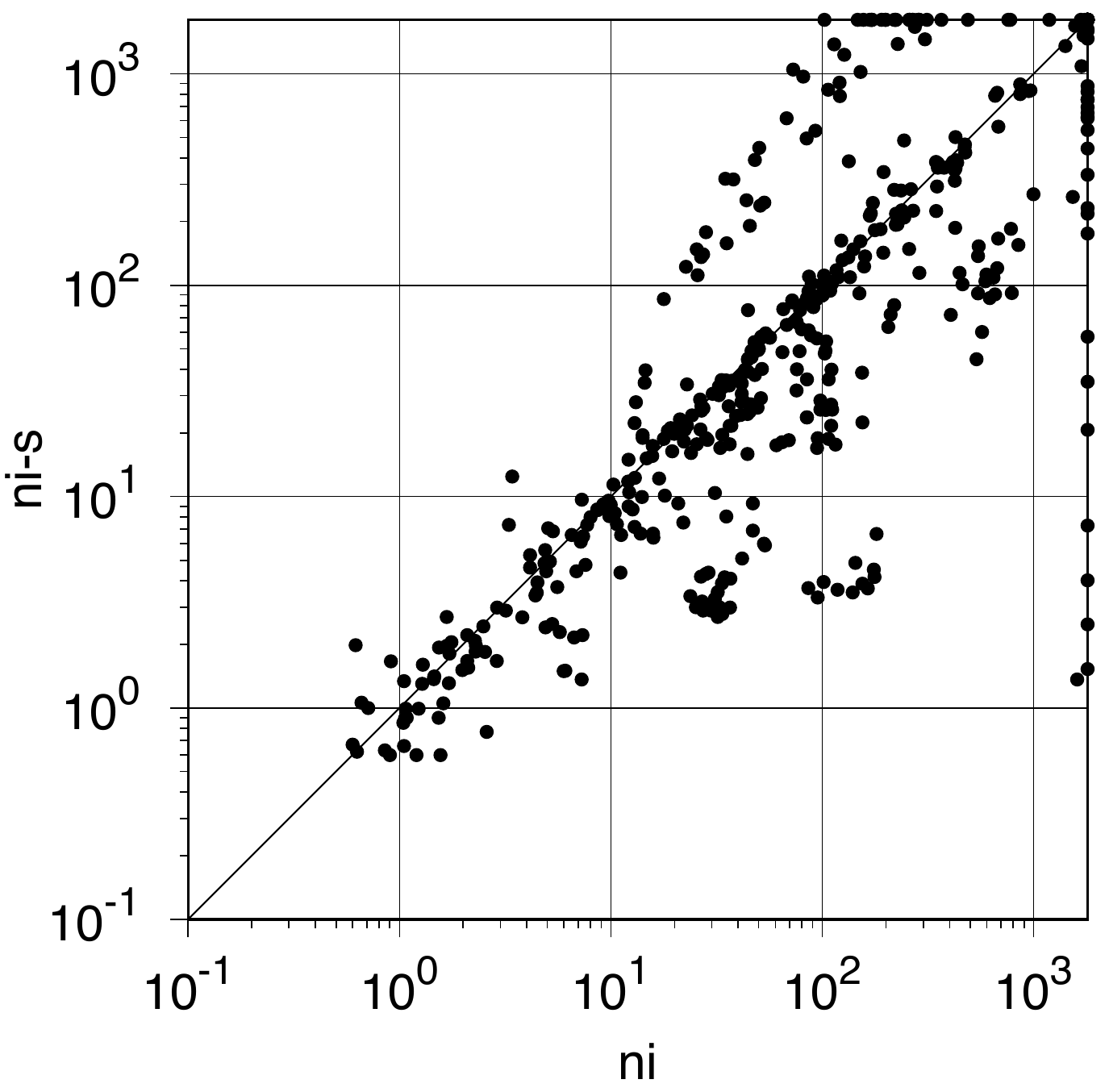} 
    \caption{With and without structure with\\ different cores} 
    \label{Fi:structure-dif-cores} 
    \vspace{4ex}
    \hfill
  \end{subfigure}%% 
  \begin{subfigure}[b]{0.5\linewidth}
    \centering
    \includegraphics[width=0.75\linewidth]{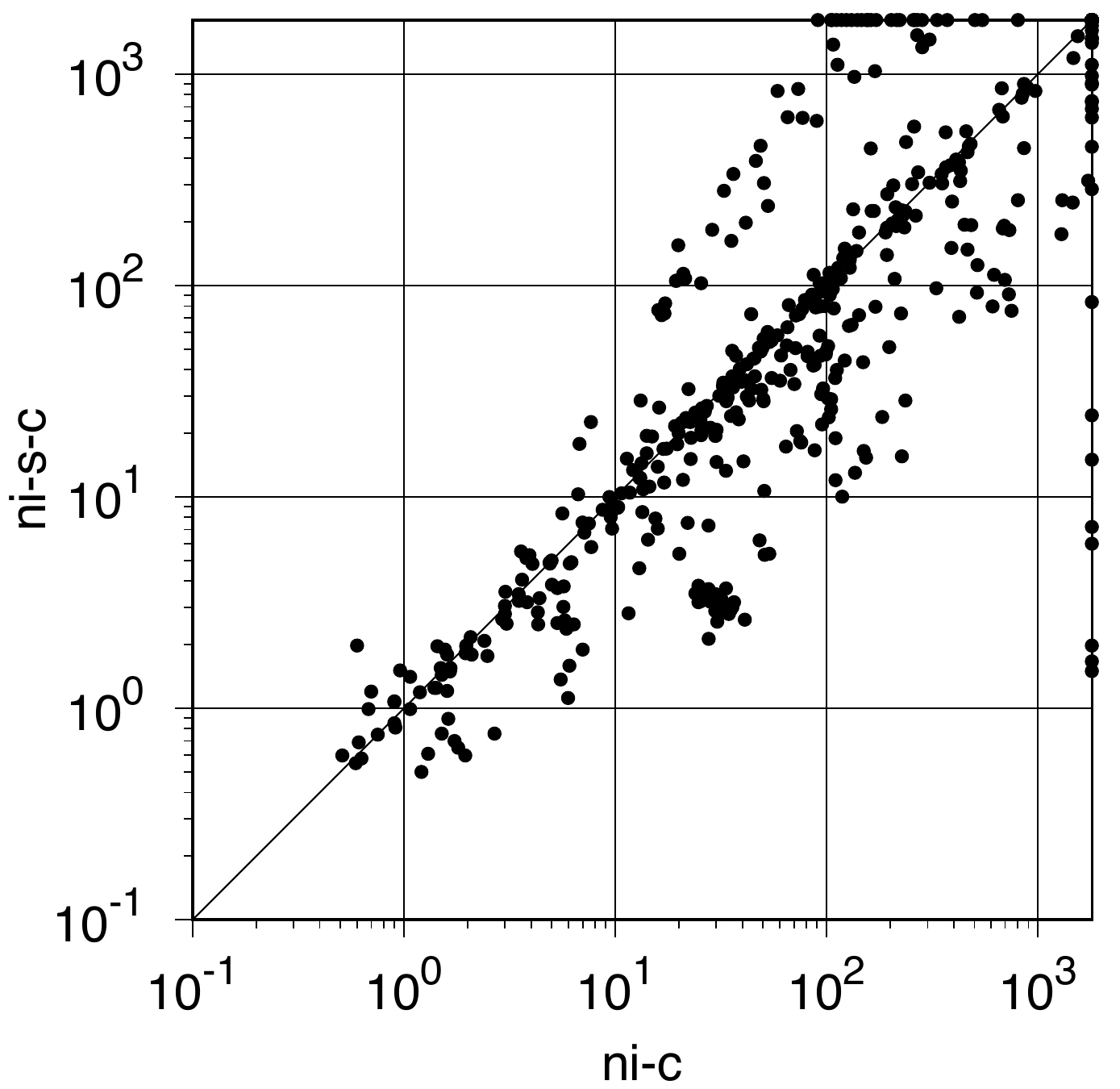} 
    \caption{With and without structure with\\ the same cores} 
    \label{Fi:structure-same-cores} 
    \vspace{4ex}
    \hfill
  \end{subfigure} 
  \vspace{-10mm}
  \caption{Impact of the structure of the encoding on the performance of the solver}
  \label{Fi:structure} 
\end{figure}

\figref{structure} shows scatter plots that compare the non-incremental versions with and without using the structure of the cores when building the cardinality encoding. Each point in the plot corresponds to a problem instance, where the x-axis corresponds to the run time required by the non-incremental versions that do not use the structure and the y-axis corresponds to the run time required by the non-incremental versions that use core structure when building the cardinality encoding. The (ni) and (ni-s) versions solve 527 and 524 instances, respectively. However, when looking at the performance of the algorithm, we can see that the majority of the instances are solved faster when using the structure of the cores to build the encoding. Nevertheless, there are several outliers where this behavior is not observed. The (ni-c) and (ni-s-c) versions solve 521 and 522 instances, respectively. These versions were run on the 555 instances solved by the incremental version and reuse the same cores found by the incremental version. Therefore, the only difference between (ni-c) and (ni-s-c) is on how the encoding is built. In this case, we can also observe a similar scenario as before where (ni-s-c) is faster on the majority of the instances. Even though the structure of the cores when building the cardinality encoding has a minor effect on the number of solved instances, it does seem to improve the running time of the solver.

\vspace{2mm}
\noindent {\bf Threats to validity.} Even though our results support our original insights that incrementality is the main reason for the observed improvement, some factors may have led us to wrong conclusions. In particular, maybe if we used a different set of benchmarks, then the results would be different. However, we did a preliminary study with different benchmarks~\cite{pai-poster} and obtained similar results. Another threat to our conclusions is the fact that we only tested our approach using the MSU3 algorithm. It may be that the structure of the cores when building the cardinality encoding is more important for other MaxSAT algorithms (e.g., WPM3~\cite{DBLP:journals/ai/AnsoteguiG17}, RC2~\cite{ignatiev2018pysat}) than for MSU3. As future work, we plan to investigate this impact on different algorithms and see if our conclusions still hold.

%% file: impact.tex
\section{Impact on MaxSAT and beyond\label{Se:impact}}

\begin{figure}
\centering
\includegraphics[scale=0.85]{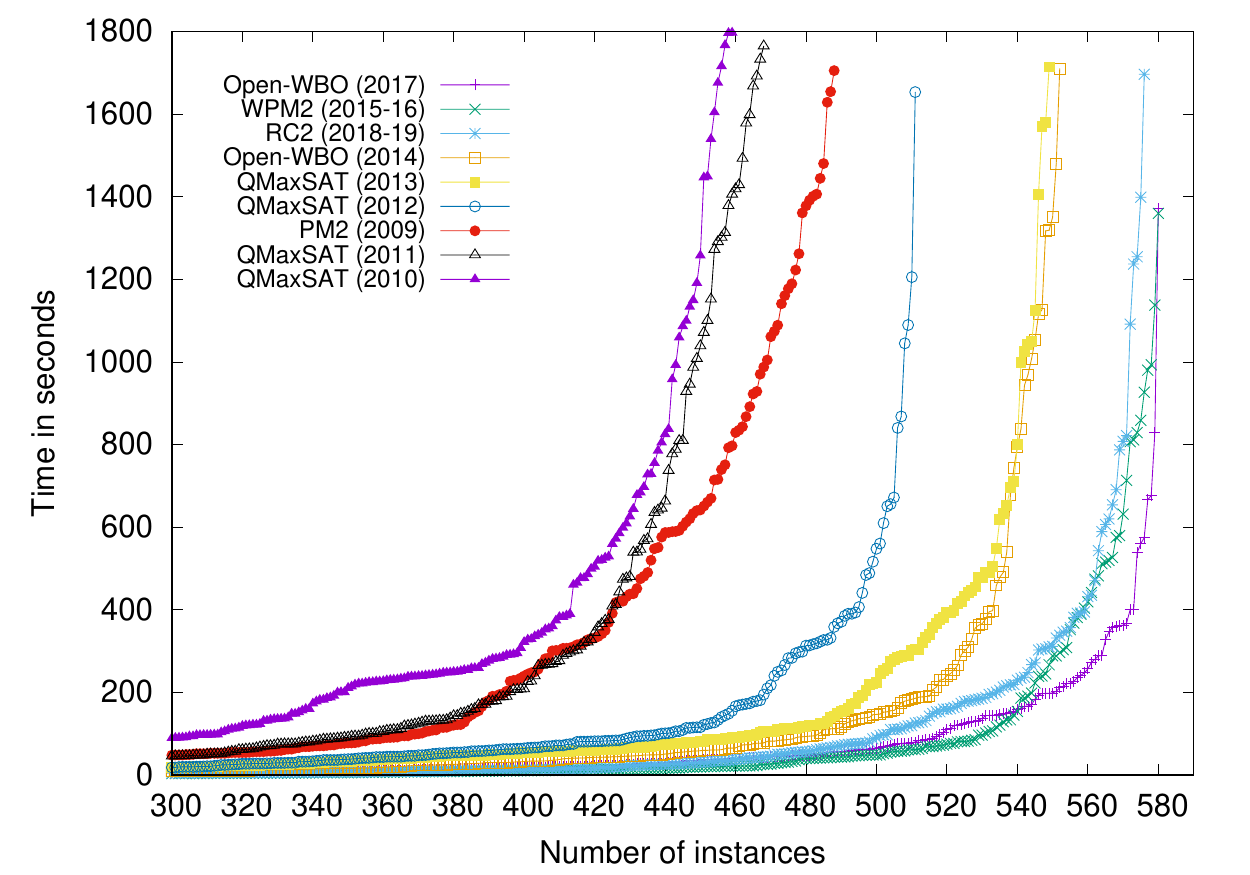}
\caption{Evolution of MaxSAT solvers for partial MaxSAT}\label{Fi:evolution}
\end{figure}

\figref{evolution} shows a cactus plot with the evolution of MaxSAT solvers for partial MaxSAT since 2009 until 2019. We selected the non-portfolio solvers with most instances solved in the MaxSAT Evaluation of each year. All solvers were run on StarExec with a time limit of 1,800 seconds and a memory limit of 32GB on the 627 partial industrial benchmarks from MaxSAT Evaluation 2013. The solvers in \figref{evolution} are: PM2~\cite{DBLP:conf/ccia/AnsoteguiBL09} (2009), QMaxSAT~\cite{qmaxsat-jsat12,totalizer-ictai13} (2010-2013), Open-WBO~\cite{DBLP:conf/cp/MartinsJML14,neves-sat15} (2014, 2017), WPM3~\cite{DBLP:journals/ai/AnsoteguiG17} (2015-16), and RC2~\cite{ignatiev2018pysat} (2018-19). We can see a significant improvement of MaxSAT solvers in the last decade.

The techniques proposed in the CP'14 paper were implemented
on top of the {\tt open-wbo} framework~\cite{martins-sat14} and won the partial industrial MaxSAT Evaluation 2014 category. 
This framework is open-source where any minisat-like SAT solver can be
pluged-in. Since the proposed techniques became available in an
open-source solver, it became easier to be adopted by the research
community.
As a result, other state of the art MaxSAT solvers have also adopted
many of the techniques, in particular the iterative encoding of
cardinality constraints~\cite{morgado2014mscg,nadel2018solving}.
In 2015, it was shown that the iterative encoding could also
be used for encoding pseudo-Boolean constraints into SAT~\cite{martins2015using}.
Nowadays, top performing Unsat-Sat MaxSAT algorithms use at
least one of the three techniques proposed in the paper either
for cardinality of for pseudo-Boolean
constraints~\cite{neves-sat15,ansotegui2016exploiting,ignatiev2018pysat}.
Moreover, a new generation of incomplete solvers for MaxSAT
also take advantage of algorithms using these
techniques~\cite{linsbps2018,satlike-c2018,guerreiro2019constraint}.

The success of solving MaxSAT using incremental encodings of constraints
has provided a boost in performance when solving problem instances
in several domains, such as program repair~\cite{fm15}, model-based
diagnosis~\cite{marques2015efficient, liu2018efficient} or timetabling~\cite{demirovic2017maxsat}.
Moreover, these ideas have also provided inspiration for similar incremental
approaches in solving other problems such as Markov logic networks~\cite{si2016incremental} or
software analysis~\cite{si2017maximum}.